\documentclass[pra,preprint,tightenlines,showpacs,twocolumn,10pt]{revtex4} 
\usepackage{epsfig} 

\begin{document} 

\title{ Relativistic time dilatation and the spectrum of electrons emitted \\[0.2cm]   
by $33$ TeV lead ions penetrating thin foils  } 

\author{ B.Najjari, A.Surzhykov and A.B.Voitkiv }   
\affiliation{ Max-Planck-Institut f\"ur Kernphysik, 
Saupfercheckweg 1, D-69117 Heidelberg, Germany } 


\begin{abstract} 

We study the energy distribution 
of ultrarelativistic electrons produced 
when a beam of $33$ TeV Pb$^{81+}$(1s) ions 
penetrates a thin Al foil. We show that, because 
of a prominent role of the excitations 
of the ions inside the foil which becomes possible 
due to the relativistic time dilatation, 
the width of this distribution can be much narrower 
compared to the case when the ions interact with 
rarefied gaseous targets. We also show that 
a very similar shape of the energy distribution 
may arise when $33$ TeV Pb$^{82+}$ 
ions penetrate a thin Au foil.  
These results shed some light on 
the origin of the very narrow electron 
energy distributions observed experimentally 
about a decade ago. 

\end{abstract} 

\pacs{34.10+x, 34.50.Fa} 

\maketitle 


Atomic physics normally does not deal 
with objects exposed to extreme conditions. 
One of the interesting and important exceptions 
of this rule is represented by the studies 
of various phenomena accompanying the penetration of 
targets by highly charged projectile ions 
moving with velocities very close to the speed of light. 
During the interaction between the ion and a target 
atom both of these particles  
are exposed to extremely intense and extraordinarily  
short pulses of the electromagnetic fields. 

For instance, in collisions of $33$ TeV hydrogen-like 
Pb$^{81+}$(1s) ions with Al (which will be considered below) 
the typical durations of the electromagnetic pulses acting on 
the electron bound in the ion are $\stackrel{<}{\sim} 10^{-21}$ 
sec (in the rest frame of the ion). The peak pulse intensities   
in this frame can reach $\sim 10^{28}$-$10^{29}$ W/cm$^2$ 
which enables, despite the very short interaction time, 
to induce transitions of the very tightly bound electron 
of the ion with a noticeable probability \cite{foot-1}.  

First experimental results on the total cross section 
for the electron loss from $33$ TeV Pb$^{81+}$(1s)  
were reported in \cite{exp-total} together with data 
for the electron capture by $33$ TeV bare Pb$^{82+}$ 
ions \cite{foot-2}.  

Compared to the study of the total cross sections 
much more information can be obtained when 
differential cross sections are explored.  
The first experimental results on the differential cross sections 
for such collisions were reported in \cite{rel-cusp-exp}.  
In that experiment the incident beams of 
$33$ TeV Pb$^{81+}$(1s) and $33$ TeV Pb$^{82+}$ 
were penetrating Al and Au foils, 
respectively. In both cases it was found that 
the penetration is accompanied by the emission 
of ultrarelativistic electrons whose energy 
distributions have the form of a cusp with 
a maximum at an energy corresponding 
to the electrons moving 
in the laboratory frame with velocities equal 
to that of the ions. 

One of unexpected results reported in 
\cite{rel-cusp-exp} was that the measured distribution 
of the high-energy electrons produced under the bombardment 
of a thin Al foil was found to be much narrower 
than one could expect based on the consideration of the width of 
the Compton profile of the electron state in 
the incident Pb$^{81+}$(1s) ions \cite{rel-cusp-exp}.  
Moreover, in a more rigorous calculation performed 
in \cite{VG-cusp} for the energy spectrum of electrons 
emitted from a $33$ TeV Pb$^{81+}$(1s) ion colliding 
with an Al atom it was confirmed  
that such a spectrum is indeed much broader 
than that observed in the experiment \cite{rel-cusp-exp}. 
 
Another intriguing finding of \cite{rel-cusp-exp} 
was that for $33$ TeV Pb$^{82+}$ ions incident 
on a thin Au foil the shape of the measured 
energy distributions of high-energy electrons 
emerged from the foil was {\bf very similar} to that 
obtained for the beam of $33$ TeV Pb$^{81+}$(1s) 
ions incident on the Al foil. 

It is known that the total and differential 
loss cross sections depend on a bound state from which 
the electron leaves the ion 
(see e.g. \cite{SDR}, \cite{geiss} and references therein).   
Therefore, it was speculated in \cite{rel-cusp-exp}  
that in the case of the incident 
$33$ TeV Pb$^{82+}$ ions the very narrow shape  
of the electron cusp might be a signature 
of the electron capture into excited states. 
However, for the Pb$^{81}$(1s) ions incident 
on the Al foil the possible influence 
of excited states of these ions   
on the electron cusp was not 
considered seriously because of 
the common experience that excitations  
of very heavy hydrogen-like ions 
inside thin foils of relatively 
light elements do not have a noticeable 
impact on the electron loss process. 

For instance, in the recent experimental-theoretical  
study \cite{geissel} on $200$ MeV/u Ni$^{27+}$(1s) 
ions incident on gaseous and solid targets 
it was found that the fraction of 
the ions excited inside the solids does not exceed 
$5$-$6 \%$. Moreover, even such rather 
modest values seem to be hardly reachable 
for very heavy hydrogen-like ions since, 
compared to the case of relatively light ions, 
the penetration of matter by the very heavy ions 
possesses the following two important differences. 

First, because of a very tight binding of  
the electron in such ions cross sections 
for collision-induced electron transitions 
are much smaller. Therefore,  
for highly charged ions, 
like Pb$^{81+}$, moving inside solids  
the mean free path with respect to 
the collision-induced transitions 
will be much larger. Second, 
the lifetimes of the excited states 
with respect to the spontaneous 
radiative decay in such ions 
are much shorter. 

The above two points mean 
that there will be much more time 
between the consequent collisions  
for the excited ion to relax into 
the ground state via the spontaneous 
radiative decay. As a result, 
there might seem to be sound grounds 
for the sceptic attitude to 
the possible role played by 
the excitations of the incident 
$33$ TeV Pb$^{81}$(1s) ions in the formation 
of the electron cusp. However, it will 
be demonstrated below that the expectations 
based on the experience accumulated 
when exploring collisions at moderate 
relativistic impact energies have to 
be substantially corrected in the case 
of the extreme relativistic 
energies studied in \cite{rel-cusp-exp}. 

Our consideration of the energy spectrum of 
the cusp electrons assumes that the foil materials 
are amorphous (not crystals) 
and includes three main ingredients. 

First, the basis of our consideration is 
represented by calculations of cross sections 
for the projectile-electron excitation (de-excitation)  
and loss occurring in the ion-atom collisions. 
Besides, we calculate also cross sections 
for the bound-free pair production (at the extreme 
relativistic collision energy which we consider 
the radiative and kinematic capture channels 
can be ignored). 
In these our calculations we use the Dirac-Coulomb 
wave functions to describe bound and continuum 
states of the electron (and the positron) 
in the field of the bare lead nucleus. 
We shall not elaborate further on the  
details of the cross section calculations 
and just refer to \cite{ABV} for the description of 
the calculation methods.  
In addition, within our basic atomic physics analysis 
we also calculate rates for the spontaneous 
radiative decay of excited hydrogen-like lead ions 
to all possible internal states with lower energies. 

In the second step we solve the kinetic equations 
which describe the population of the internal states 
of the ion inside the foil given as a function of time $t$ 
or of the ionic coordinate $z=vt$ inside the foil   
($z$ and $t$ are measured in the laboratory frame, 
${\bf v}=(0,0,v)$ is the projectile velocity). 
These equations read   
\begin{eqnarray}
\frac{d N_0}{d t} &=& - \frac{N_0}{\tau^{capt}} %
+ \sum_{j=1}^{N_{max}} \frac{N_j}{\tau_j^{loss}},  %
\nonumber \\     
\frac{d N_j}{d t} &=& \frac{ N_0 }{ \tau_j^{capt} } %
- \frac{ N_j }{ \tau_j^{loss} } 
- N_j \sum_{i=1}^{i \leq j } \frac{ 1 }{ \tau^{sp}_{j \to i} } %
\nonumber \\     
&& - N_j \sum_{i=1 (i \neq j)}^{N_{max}} \frac{ 1 }{ \tau_{j \to i} } %
+ \sum_{i=1 (i \neq j)}^{N_{max}} \frac{ N_i }{ \tau_{i \to j} }. 
\label{e1}
\end{eqnarray}
Here, $N_0$ is the number of bare ions,   
$N_j$ is the number of ions with one electron in the $j$-th 
internal state ($j=1,2,..,N_{max}$) and $N_{max}$ 
is the total number of the involved bound states. 
Further, $\tau_j^{capt}$ is the mean time 
for the electron vacuum capture into the $j$-th state,    
$\tau^{capt}$ is the mean time for the electron capture 
to any state ($ 1/\tau^{capt}=1/\tau_1^{capt}+1/\tau_2^{capt}+...$), 
$\tau_j^{loss}$ is the mean time for the electron loss 
from a state $j$ to the continuum, 
$\tau_{j \to i}$ is the mean time for the collision 
induced transition from the internal state $i$ to the internal 
state $j$ and $\tau^{sp}_{j \to i}$ is the lifetime 
of the state $j$ with respect to the spontaneous radiative 
transition to any possible state $i$. 

The elementary cross sections and spontaneous 
decay rates obtained during the first step of 
the consideration enable one to get the above 
mean excitation/de-excitation loss and 
capture times in the usual way. For instance, 
$\tau_j^{loss}=1/(n_a \sigma_j^{loss} v)$, where 
$\sigma_j^{loss}$ is the cross section for 
the electron loss from the $j$-th internal state 
of the ion and $n_a$ is the atomic density of the target. 
 
Once the functions $N_j(z)$ are known, 
the (preliminary) estimate for the 
energy spectrum of the electrons 
emitted from the ion traversing a solid foil 
of a thickness $L$ is given by 
\begin{eqnarray}
\frac{ d n_e }{ d \varepsilon_p } = %
n_a \sum_{j=1}^{ N_{max} } %
\frac{ d \sigma_j^{loss} }{ d \varepsilon_p } %
\int_0^L dz N_j(z),   
\label{e2} 
\end{eqnarray}
where $ \varepsilon_p $ is the total electron energy in the 
laboratory frame and $\frac{ d \sigma_j^{loss} }{ d \varepsilon_p }$
is the energy distribution of the electrons emitted from 
the internal state $j$. 

The third step of our consideration deals with 
the transport of the emitted electrons through the foil.   
The detailed analysis of this step represents in general 
quite a delicate task but in our case is substantially 
simplified by the fact that the electrons leaving the ions 
have in the laboratory frame extremely high values of energy.
The are two main effects which can influence the shape of 
the electron energy distribution when the electrons 
penetrate the foil. 

The first concerns energy losses of the ultrarelativistic 
electrons traversing the foil. These losses are caused by  
(i) the excitation of the electrons of the foil and 
(ii) the emission of the radiation by the ultrarelativistic 
electrons because of their acceleration during the interactions 
with the atomic nuclei in the foil. However, for the foil 
parameters used in the experiment \cite{rel-cusp-exp} 
the energy losses can simply be ignored because they 
are very small $ \left( \stackrel{<}{\sim} 0.5 \%  \right) $  
compared to the initial energies of these electrons. 

The second effect which may possibly influence 
the shape of the measured energy distributions is that  
collisions in the foil broaden 
the distribution of the ultrarelativistic electrons 
over the transverse components $(p_x,p_y)$ of their momenta. 
For the foil parameters used in \cite{rel-cusp-exp},  
the multiple collisions suffered by 
the ultrarelativistic electrons inside 
the foil substantially increase the width of their   
$(p_x,p_y)$-distribution compared to 
that which these electrons have when leaving 
the $33$ TeV nuclei. 

Nevertheless, even after this increase 
the transverse components ($\sim 10^2$ a.u.) remain very small 
compared to the total electron momenta ($\simeq 2 \times 10^4$ a.u.) 
which means that the broadening of the $(p_x,p_y)$-distribution
may have an impact on the measured electron momentum 
distribution only if special geometric conditions are employed 
in an experiment \cite{f-new-new}.  
Since we do not possess all necessary information 
about the real conditions of the experiment \cite{rel-cusp-exp}, 
in our calculations for the energy spectra, discussed below, 
we simply take all electrons (whichever angle they 
have after leaving the foil) into account.  

Under such conditions the changes in the electron momenta 
during the electron transport through the foil 
do not have an impact on the final electron energy distribution.   
Therefore, the main essential difference between 
the previous estimates for the shape of the electron cusp 
in the case of $33$ Pb$^{81+}$(1s) incident on an Al foil 
(where it was assumed that the electron loss  
occurs in the single collision regime) and our present model 
is that the latter takes into account electron transitions 
to the continuum not only directly 
from the ground state of the ions but also via 
the intermediate excitations to higher bound states 
occurring when the ions penetrate the foil. 

\begin{figure}[t]
\vspace{-0.35cm}
\begin{center}
\includegraphics[width=0.47\textwidth]{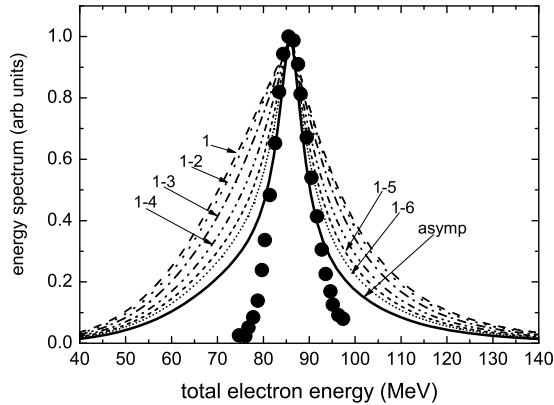}
\end{center} 
\vspace{-0.9cm} 
\caption{ The energy distribution of the electron cusp 
produced in collisions between an incident beam 
of $33$ TeV Pb$^{81+}(1s)$ with Al foil with 
a thickness of $2.85 \times 10^{-2}$ cm 
(for more explanation see the text).  
Circles show the electron energy distribution 
measured in \cite{rel-cusp-exp} for $33$ TeV Pb$^{81+}$ 
colliding with the Al foil of the same thickness.    
All the distributions are given in the laboratory frame 
and are normalized to $1$ at the maximum. } 
\label{cusp-1} 
\end{figure} 

The initial expectations, that in the case of very 
heavy ions their excitations are of minor importance 
for the loss process, seems to be just confirmed 
if we compare in figure \ref{cusp-1} curves labeled 
with '$1$' and '$1$-$2$'. 
In this figure, where we present results for the electron energy 
spectrum in the case of $33$ Pb$^{81+}$(1s) ions incident 
on Al foil, the curve '$1$' was obtained 
by ignoring all excited bound states 
while in the calculation resulted in the curve 
'$1$-$2$' the states with the principal quantum number $n=2$ 
were also taken into account. Yet, there is just 
a tiny difference in the widths of these two curves.  

However, when we add the states with $ n=3 $ 
into our analysis (the curve in figure \ref{cusp-1} 
labeled by '$1$-$3$' ) the width-reducing effect  
becomes quite visible. Adding into the analysis 
the states with $n=4$ leads to a further reduction 
in the calculated width and this reduction is even larger  
than that observed when the states with $n=3$ were added.  
The reduction of the width continues further 
when we add states with $n=5$ and $n=6$ 
(see figure \ref{cusp-1}), however, it  
proceeds at a smaller pace compared to that when 
the states with $n=3$ and $n=4$ were added. 

Note that the inclusion of the states with 
$n=1$-$6$ into the analysis means that we calculated  
the collision-induced and spontaneous radiative 
transitions in the system of levels involving 
$182$ quantum bound  states of the Pb$^{81+}$ ion as well 
as the electron and positron continua in the field 
of the nucleus Pb$^{82+}$. 
This is quite demanding and computationally expensive task.   
Due to obvious reasons in our calculations we cannot  
increase indefinitely the number of bound states.   
Therefore, we have applied an extrapolation procedure 
in order to get the asymptotic limit for 
the electron cusp shape effectively corresponding 
to taking into account all bound states 
($n=1$-$\infty $). The result of this extrapolation 
is shown in figure \ref{cusp-1} by 
the curve labeled '{\it asymp}'. 

\begin{figure}[t]
\vspace{-0.35cm}
\begin{center}
\includegraphics[width=0.47\textwidth]{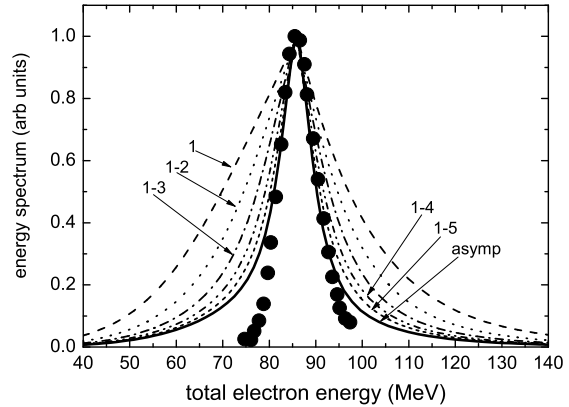}
\end{center} 
\vspace{-0.9cm} 
\caption{ Same as in figure \ref{cusp-1} but for 
an incident beam of $33$ TeV Pb$^{82+}$ penetrating 
Au foil with a thickness of $8.81 \times 10^{-4}$ cm 
corresponding to the conditions of the capture experiment 
\cite{rel-cusp-exp}. For more information, 
circles show the electron spectrum measured 
in \cite{rel-cusp-exp} for $33$ TeV Pb$^{81+}$(1s) 
ions incident on the Al foil. } 
\label{cusp-2} 
\end{figure}

Comparing the energy distributions in figure \ref{cusp-1} 
we see that their asymptotic width is about a factor 
of $3$ smaller than the width obtained by assuming 
that the cusp is produced under the single-collision conditions.  
This strong effect is caused by the excitation of the 
ions inside the foil which involves rather 
highly lying bound states: when the ions move 
in the foil the electron cloud    
surrounding the ionic nuclei has  
enough time to expand tremendously in size    
before it will almost completely disappear 
due to the transitions to the continuum. 
The key factor making this possible is the relativistic 
time dilatation which effectively decreases the spontaneous 
decay rates of the excited states of the ions by 
a factor of $\approx 170$. 

One more point which should be mentioned 
is that cross sections for the electron capture 
are relatively very small. As a result, 
in the formation of the electron cusp in the case of 
the hydrogen-like ions incident on the Al foil 
the capture channels do not play any noticeable role.

In figure \ref{cusp-2} we show results for 
the energy spectrum calculated for 
$33$ TeV Pb$^{82+}$ incident on a Au foil. 
Of course, now the electron capture from vacuum 
becomes of paramount importance for the very existence 
of the electron cusp. One should note, however, 
that the capture cross sections decrease very rapidly when 
$n$ and $j_e$ increase ($j_e$ is the total angular 
momentum of the electron in a bound state). 
Therefore, the most of the excited bound states 
having a very important impact on the energy spectrum 
are populated not by capturing the electron directly 
from the vacuum but via the excitations 
from few states with the lowest values of $n$ and $j_e$ 
for which the capture is efficient. This indirect way 
becomes especially effective because in collisions 
with Au atoms the excitation cross sections 
are much larger than in the case with Al.  
   
Comparing the spectra shown in figure \ref{cusp-2} 
with those displayed in figure \ref{cusp-1} 
we see that the changes in the form of 
the calculated spectrum in figure \ref{cusp-2} 
(occurring when we allow for more bound states 
in our analysis) are accumulating at a different pace. 
Besides, the asymptotic cusp shape 
in figure \ref{cusp-2} has less pronounced wings.  
These differences are related to two basic reasons: 
(i) the excitation/loss cross sections  
in an Au foil are much larger while the spontaneous 
decay rates remain exactly the same as 
in the case of an Al foil and 
(ii) the initial step in the cusp formation 
is now represented by the capture process 
which also somewhat increases the 
relative population of the excited states 
compared to the case when the beam of 
Pb$^{81+}(1s)$ ions was incident  
on the Al foil.  

Curiously, however, that the asymptotic width 
in figure \ref{cusp-2} is again about $3$ times 
smaller than the 'initial' width and the shape of  
the asymptotic spectra in both cases looks 
similar (which is also in agreement with 
the experimental observations of \cite{rel-cusp-exp}).  
In general such a similarity will not hold 
when the foil parameters (for instance, their thicknesses) 
are changed and, in this sense, is accidental. 
Yet, in both cases the strong reductions 
in the widths of the energy distributions are caused by  
the excitations of the electrons to rather 
high lying bound ionic states occurring 
when the ions penetrate the foils.  

In summary, briefly, we have considered 
the energy spectra of the ultrarelativistic electrons 
emitted when incident $33$ TeV Pb$^{81+}$(1s) 
and Pb$^{82+}$ ions penetrate Al and Au foils, 
respectively. The foil thicknesses were taken 
to be the same as used in the experiment \cite{rel-cusp-exp}. 
We have found that these spectra 
are much narrower than those which would 
be produced under the single-collision conditions 
and have similar shapes. The similarity in the shapes 
in general will not hold if foils with other parameters 
would be used and, thus, is fortuitous. 
However, in both cases the strong width reduction is caused 
by the excitations of the ions when they penetrate 
the target foils suffering multiple collisions with 
the target atoms. Such a profound role of the excitations 
in the case of very heavy ions is in contrast to 
the previous experience gained when exploring 
collisions in the low and intermediate relativistic 
domains of impact energies. In the case under consideration 
the excitations become so effective 
because of the relativistic time dilatation which 
decreases very strongly the spontaneous decay rates 
of excited states in the ions moving with velocities 
closely approaching the speed of light.  
   
Our present results shed some light on the origin of the 
unexpectedly narrow shape of the electron cusp 
produced by the ultrarelativistic heavy ions. However, 
a more careful analysis taking into account 
all real conditions of the experiment \cite{rel-cusp-exp}  
would be necessary in order to make a detailed comparison 
between the experiment and theory.


\begin{thebibliography}{99}

\bibitem{foot-1} These field parameters 
may be compared with the parameters of state 
of the art laser systems whose shortest 
pulse lengths are about 
$\sim 10^{-16}$-$10^{-15}$ sec and 
whose peak intensities do not exceed 
$\sim 10^{21}$-$10^{22}$ W/cm$^2$.   

\bibitem{exp-total} H.F. Krause et al, 
Phys. Rev. Lett. {\bf 80}, 1190 (1998).   

\bibitem{foot-2} Note that at such impact energies 
the capture proceeds via the electron-positron pair 
production in which the electron is created in 
a bound state of the ion, most likely in the ground state 
($\approx 75$-$80\%$ of the total capture).  

\bibitem{rel-cusp-exp} C.R. Vane et al, in 
{\it The Physics of Electronic and Atomic Collisions} 
edited by Y.Itikawa et al, p. 709 (2000)
(Proceedings of the XXI ICPEAC, Sendai, Japan, 1999).   

\bibitem{VG-cusp} A.B. Voitkiv and N. Gr\"un, 
J.Phys. {\bf B 34} 267 (2001).  
 
\bibitem{SDR}  N. Stolterfoht, R.D. DuBois and
R.D. Rivarola, {\it Electron Emission
in Heavy Ion-Atom Collisions } (Springer, Berlin, 1997). 

\bibitem{geiss} H.Geissel et al, Nucl. Instr. Meth. 
{\bf B 195}, 3 (2002).    

\bibitem{geissel} H.Ogawa et al, Phys. Rev. {\bf A 75},  
020703 (2007).   

\bibitem{ABV} A.B. Voitkiv, Phys. Rep. {\bf 392}, 191 (2004). 

\bibitem{f-new-new} For instance, 
if an experiment detects only those high-energy 
electrons, which finally move inside a very narrow cone centered 
along ${\bf v}$, the expansion the electron $(p_x,p_y)$-distribution 
inside the foil will first of all impact the detected 
numbers of the electrons emitted from the ground state 
of the lead ions (both in absolute and relative proportions) 
because these electrons right after the emission from the ions  
have larger transverse momenta (and, besides, are the first 
to appear in the continuum). Compared to the emission from 
excited states the electrons ejected from the ground state 
have the largest width of the energy distribution. Therefore, 
a comparatively stronger removal of those electrons, 
which were emitted from the ground state, 
occurring when the electron beam traverses the foil 
will effectively decrease the width of the measured 
electron energy distribution. 

\end{thebibliography}
\end{document}